\documentclass[utf8]{FrontiersinVancouver}

\setcitestyle{square} % for articles in the journals "Frontiers in Physics" and "Frontiers in Applied Mathematics and Statistics" 
\usepackage{url,hyperref,lineno,microtype,subcaption}
\usepackage[onehalfspacing]{setspace}

\usepackage{doi}    % makes printed DOIs clickable
\hypersetup{
  colorlinks=true,
  linkcolor={red!60!black},
  anchorcolor={green!80!black},
  urlcolor={blue!80!black},
  citecolor={blue!50!black}
}

\def\keyFont{\fontsize{8}{11}\helveticabold }
\def\firstAuthorLast{Hu} %use et al only if is more than 1 author
\def\Authors{Bai-Shan Hu\,$^{1,*}$}
\def\Address{$^{1}$School of Physics, and State Key Laboratory of Nuclear Physics and Technology, Peking University, Beijing 100871, China}
\def\corrAuthor{Bai-Shan Hu}

\def\corrEmail{baishanhu@pku.edu.cn}

\begin{document}
\onecolumn
\firstpage{1}

\title[Ab Initio Nuclear Theory for Heavy Nuclei and Its Application to Dark Matter-Nucleus Scattering]{Ab Initio Nuclear Theory for Heavy Nuclei and Its Application to Dark Matter-Nucleus Scattering} 

\author[\firstAuthorLast ]{\Authors} %This field will be automatically populated
\address{} %This field will be automatically populated
\correspondance{} %This field will be automatically populated

\extraAuth{}% If there are more than 1 corresponding author, comment this line and uncomment the next one.
%\extraAuth{corresponding Author2 \\ Laboratory X2, Institute X2, Department X2, Organization X2, Street X2, City X2 , State XX2 (only USA, Canada and Australia), Zip Code2, X2 Country X2, email2@uni2.edu}

\maketitle

\begin{abstract}

%%% Leave the Abstract empty if your article does not require one, please see the Summary Table for full details.
\section{}

The era of precision ab initio nuclear theory has arrived, enabling uncertainty-quantified predictions for nuclear structure and for interactions with external probes directly from the underlying nuclear force and electroweak currents. This review highlights recent breakthroughs that extend ab initio calculations to the heavy nucleus $^{208}$Pb, to medium-mass systems with complex deformation, and to weakly-bound nuclei near the driplines. We also summarize ab initio calculations of nuclear responses for dark matter direct detection. Together, these advances demonstrate how ab initio methods can substantially reduce nuclear-physics uncertainties in searches for physics beyond the Standard Model, providing a more robust interpretation of current and forthcoming precision experiments.

\tiny
 \keyFont{ \section{Keywords:} ab initio nuclear theory, effective field theory, beyond the Standard Model, dark matter detection, nuclear form factors, collective correlations, continuum coupling} %All article types: you may provide up to 8 keywords; at least 5 are mandatory.
\end{abstract}

\section{Introduction}

Ab initio nuclear theory seeks to solve the nuclear many-body problem directly from the underlying nuclear forces and electroweak currents. The approach starts with realistic nuclear forces that reproduce nucleon-nucleon scattering data and observables of few-body systems (mass number $A\le4$) with high precision. In particular, two- and three-body nuclear forces derived from effective field theory (EFT), along with consistent electroweak currents, have become the ``standard" inputs. The resulting many-body Schr\"odinger equation is then solved using systematically improvable many-body solvers, which provide approximation-controlled results without any phenomenological adjustments. Over the past decade, ab initio nuclear theory has made rapid progress in the range of masses and physics that can be treated~\cite{hergert2020,Ekstrom2023fphy,Machleidt2023,hu2022a,hebeler2023,ye2025}. This paradigm shift allows us to address some of the most exciting questions in nuclear structure, nuclear astrophysics, and the frontiers of physics beyond the Standard Model (BSM), starting solely from the underlying nuclear and electroweak forces. 

A new generation of cutting-edge, large-scale facilities, such as FRIB (USA), RIBF (Japan), FAIR (Germany), RAON (South Korea), and HIAF (China), are either in the commissioning phase or set to become operational in the coming years \cite{ye2025}. These powerful tools, in synergy with high-precision measurements near the valley of stability, are poised to explore the \emph{terra incognita} of the nuclear chart and generate an unprecedented wealth of experimental data. This coming data flood, such as nuclear masses, dripeline locations, low-lying spectra, electroweak transitions and moments, presents a tremendous opportunity for ab initio nuclear theory to not only explain new observations but also to make robust predictions that can guide future experiments. Such theoretical work is crucial for addressing fundamental questions regarding the nuclear forces that bind nucleons into nuclei, the formation of stable and rare isotopes, the emergence of complex collective phenomena, and their applications in nuclear astrophysics. Because ab initio descriptions inherit the physics of the Standard Model through EFT, they are expected to provide more reliable and powerful predictions compared to traditional phenomenological models.

Meanwhile, high-precision experiments targeting dark matter (DM), neutrinos, and gravitational waves have become central to probing the structure of matter and fundamental interactions. At present, dozens of large-scale experiments are underway worldwide, including DUNE, Hyper-K, JUNO, the COHERENT program, LZ, XENONnT, PandaX, CDEX, and aLIGO. In many of these frontier searches, such as DM direct detection, neutrino measurements, studies of CP violation, and observations of neutron stars and supernovae, uncertainties in the structure of target or detector nuclei are often an important limitation on experimental precision. Ref. \cite{eXdb} lists specific experiments and the nuclei involved, spanning from $^2$H to $^{208}$Pb. Most of these nuclei are located in the medium- or heavy-mass regions, such as $^{40}$Ar, $^{48}$Ca, $^{\rm nat}$Ga, $^{\rm nat}$Ge, $^{82}$Se, $^{90}$Zr, $^{100}$Mo, $^{116}$Cd, $^{128,130}$Te, $^{127}$I, $^{\rm nat}$Xe, $^{133}$Cs, $^{150}$Nd, $^{\rm nat}$Gd and $^{\rm nat}$Pb, posing significant challenges for ab initio nuclear theory. Advancing ab initio nuclear theory to describe these nuclei and their interactions with particles like DM and neutrinos is therefore essential. Such efforts provide the crucial nuclear-physics input needed for new discoveries and offer a path toward deeper insights into the fundamental properties of matter and BSM physics.

This work is organized as follows. Section 2 summarizes some recent advances in chiral nuclear forces and ab initio nuclear structure studies. Section 3 then focuses on recent progress in ab initio calculations of nuclear responses for DM direct detection.

\section{Ab Initio Nuclear Theory}

The basic model of nuclear theory assumes that a nuclear system can be described by a non-relativistic Hamiltonian that contains interactions among nucleons, i.e., protons and neutrons. An ab initio nuclear calculation starts with the intrinsic Hamiltonian of an $A$-body nuclear system, which reads
\begin{equation}
H=\sum_{i<j}^{A}\left(\frac{(\vec{p}_{i} - \vec{p}_{j})^2}{2m_{N} A} + V_{i j}^{\mathrm{NN}}\right)+\sum_{i<j<k}^{A} V_{i j k}^{3 \mathrm{N}},
\label{H_in}
\end{equation}
where $\vec{p}$ is the nucleon momentum in the laboratory, $m_{N}$ is the nucleon mass, $V^{\rm NN}$ and  $V^{\rm 3N}$ represent the two-nucleon (NN) and three-nucleon (3N) interactions, respectively. The Coulomb interaction is also included in the two-body interaction part. Once the $V^{\rm NN}$ and  $V^{\rm 3N}$ are constructed, a state-of-the-art quantum many-body method is employed to solve the many-body Schr\"odinger equation \eqref{H_in}. We will discuss these two components, the NN plus 3N interactions and quantum many-body methods, in the following sections.

\subsection{Interactions from chiral effective field
theory}

Since Steven Weinberg proposed the groundbreaking idea of applying chiral EFT to derive nuclear interactions in the early 1990’s, the continuous development of chiral EFT has revolutionized the field of nuclear physics \cite{Weinberg1990,Weinberg1991}. This framework provides a solid theoretical foundation and a systematically improvable scheme for constructing two- and many-nucleon interactions consistent with QCD's symmetries and symmetry-breaking patterns. Recent advances in chiral EFT are pushing nuclear forces to new levels of accuracy and applicability, establishing chiral interactions as the new “standard” inputs for ab initio nuclear studies.

Significant advances have been made in constructing chiral nuclear interactions over the past decade (see e.g., Refs~\cite{RevModPhys.81.1773, Machleidt2011a, RevModPhys.92.025004, Tews2022} and references therein). This review, however, focuses specifically on progress in calibrating the unknown low-energy constants (LECs). These constants parameterize the unresolved short-distance physics of the strong force and are typically constrained by experimental data, such as nucleon-nucleon scattering phase shifts and few-body (A$\leq$4) observables. As the EFT expansion order increases, the number of LECs grows rapidly, yielding a high-dimensional parameter space. This complexity makes LEC calibration particularly challenging and introduces the risks of overfitting and converging to local minima.

In recent works \cite{hu2022a,hu2024hm,elhatisari2024,Jiang2024hm}, the history matching approach has been employed to explore and reduce the vast number of different parameterizations within the high-dimensional parameter space of LECs. History matching is an iterative process that identifies and eliminates implausible parts of the input space based on an implausibility metric, shrinking the input space in each iteration, and repeating the search for non-implausible inputs in the smaller space. By confronting with data in nucleon-nucleon scattering and selected light nuclei or even heavier-mass nuclei, while accounting for relevant uncertainties, the history matching approach can identify the so-called non-implausible domain of LECs that cannot be ruled out. There is also an effort to find a single chiral interaction that can precisely reproduce bulk properties, spectra, and nuclear matter saturation properties. This chiral two- and three-nucleon interaction, called N$^3$LO$_{\rm Texas}$ \cite{Hu2025a}, was optimized using a novel optimization protocol that employs high-fidelity emulators of few- and many-body observables -- nucleon–nucleon scattering phase shifts and the properties of $^2$H, $^4$He, and $^{16}$O -- in combination with a new multi-start global optimization method. This protocol can explore a vast LEC landscape, effectively mitigating overfitting and avoiding local
minima. Uniquely, it enables the inclusion of accurate coupled-cluster calculations for $^{16}$O in the calibration, which is crucial for constraining high-partial wave LECs and achieving accurate binding energies and charge radii in medium- and heavy-mass nuclei. Furthermore, although nuclear-matter properties were not included in the optimization, the N$^3$LO$_{\rm Texas}$ describes nuclear matter near saturation point in a manner consistent with empirical constraints.

\subsection{Nuclear many-body methods}

Once the $A$-body Hamiltonian \eqref{H_in} is constructed, many-body solvers are employed to treat nuclear systems ranging from few-nucleon systems to heavy nuclei and even infinite nuclear matter~\cite{MortenBook,hergert2020,ye2025,Marino2024nm,hu2025qmc}. These methods can be broadly divided into two categories: ``virtually exact” and ``powerfully approximate” \cite{Soma2020}. Methods in the first category solve the many-body Schr\"odinger equation without any formal approximation, meaning their accuracy is limited only by basis truncation and numerical precision. Examples include Quantum Monte Carlo (QMC)~\cite{RevModPhys.87.1067,Roggero2014,Lonardoni2018,Arthuis2023} and no-core shell model (NCSM)~\cite{Barrett2013,PhysRevC.79.014308,PhysRevLett.106.202502,Liebig2016}. The second category includes approaches that approximate the solution in a systematically improvable way. Examples include lattice effective field theory (LEFT)~\cite{Lahde2019:left,lee2025,PhysRevLett.106.192501,Lu2022:pt,Ma2024:nm,elhatisari2024}, coupled cluster (CC)~\cite{RevModPhys.79.291,shavittbartlett2009,Hagen2014,PhysRevC.76.034302,PhysRevLett.113.142502,hu2024Zr,Sun2025a,p297-y8vq}, self-consistent Green's function (SCGF)~\cite{Dickhoff2004, Som2020, PhysRevC.101.014318}, in-medium similarity renormalization group (IMSRG)~\cite{PhysRevLett.106.222502,PhysRevC.85.061304,Hergert2016,stroberg2017,Hu2019gimsrg,yao2020,PhysRevC.105.L061303,Zhen2025} and many-body perturbation theory (MBPT)~\cite{ROTH2010272,PhysRevC.94.014303,PhysRevLett.122.042501,10.3389/fphy.2020.00164}. The computational cost for the ``exact" methods increases exponentially with increasing mass number, and they have achieved accurate descriptions of light nuclei. In contrast, the approximate methods have a cost that increases polynomially with increasing system size. This favorable polynomial scaling has pushed the frontier of ab initio computations from light nuclei towards heavier nuclei \cite{Morris2018Sn,Arthuis2020}, recently reaching $^{208}$Pb \cite{hu2022a,hebeler2023}. Given that most target nuclei in DM direct detection experiments lie in the medium- to heavy-mass regions and are double open-shell systems, the IMSRG method and its variants \cite{HERGERT2016165,PhysRevLett.118.032502,Hu2019gimsrg,yao2020,PhysRevC.105.L061303} provide the most prominent ab initio nuclear calculations.

\subsection{Some advances in ab initio calculations of heavy nuclei}

\begin{figure}[h!]
\begin{center}
\includegraphics[width=12cm]{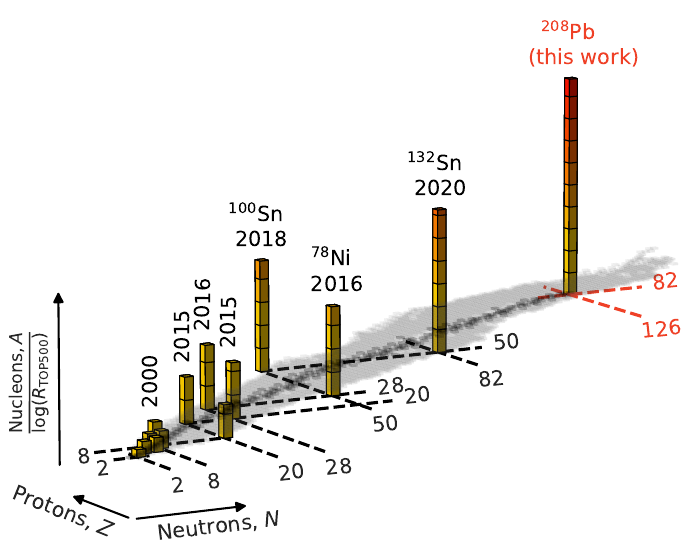}
\end{center}
\caption{Trends in ab initio calculations for the nuclear $A$-body problem, adapted from Ref.~\cite{hu2022a}. The bars mark the years of the first realistic computations of doubly magic nuclei. The height of each bar is the mass number $A$ divided by the logarithm of the total computational power R$_{\rm TOP500}$ (in flop/s) from the pertinent TOP500 list (https://www.top500.org/statistics/perfdevel/). The point labeled ‘this work’ refers to the calculation in year 2022 \cite{hu2022a}.}\label{fig:1}
\end{figure}

Figure~\ref{fig:1} shows the progress of ab initio calculations, using key doubly magic nuclei as benchmarks. The modern era of ab initio nuclear theory began around the year 2000, and after two decades of progress, the first accurate computation of the double magic nucleus $^{208}$Pb was achieved in year 2022 \cite{hu2022a}. This achievement was reached by combining advances in the Bayesian history matching approach, quantum many-body methods, and emulator technology \cite{hu2022a}. Notably, Ref.~\cite{hu2022a} also found that the neutron skin in $^{208}$Pb and the slope parameter $L$ of the nuclear matter equation of state (EoS) are strongly correlated with nucleon-nucleon scattering phase shifts in the $1S_0$ partial wave at laboratory energies around 50 MeV, offering new insights into our understanding of neutron skins and neutron stars. This study also demonstrated that it is possible to compute nuclei from A=2 to A$\approx$208, as well as infinite nuclear matter, using consistent chiral nuclear interactions. The gray background in Fig.~\ref{fig:1} illustrates the full nuclear landscape, and we can see that $^{208}$Pb is located in the heavy-mass region, and its accurate computation signifies that ab inito nuclear theory has entered a new phase of pursuing quantitative descriptions of nuclei across a wide range of mass
numbers.
%pursuing a quantitative description of the entire chart of nuclei.

Although ab initio computations have reached the $^{208}$Pb region, this nucleus has closed proton and neutron shells and, therefore, exhibits a relatively simple, spherical structure. Computing its ground state requires capturing ``only" dynamical correlations. By contrast, nuclei with strong static or collective correlations, such as strongly deformed nuclei or those exhibiting shape coexistence, remain a significant challenge \cite{hergert2020}. To address this, recently developed frameworks describe doubly open-shell nuclei by simultaneously including both dynamical and collective correlations, such as in-medium generator coordinate method (IM-GCM)~\cite{yao2020,Zhou2025,Zhou2026:N20}, CC~\cite{Hagen2022Mg,hu2024Zr,hu2024Ni,Sun2025a} IMSRG~\cite{PhysRevC.105.L061303,Cao2025}, and MBPT \cite{Frosini2022a,Frosini2022b,Frosini2022c}. These computations capture dynamical correlations via particle-hole excitations from deformed reference state(s) and collective correlations via angular momentum projection. Such studies, spanning from light nuclei recently up to the $A \approx 80$ region, have successfully reproduced the key features of rotational bands and B(E2; 0$^+$ $\rightarrow$ 2$^+$) transitions in deformed nuclei, and have also indicated the presence of shape coexistence \cite{hu2024Zr,hu2024Ni,Zhou2025,Cao2025}.

Another challenge in nuclear theory today is the description of weakly-bound and unbound isotopes located near the driplines. These exotic nuclei are an important focus of experimental programs at current and next-generation rare-isotope-beam (RIB) facilities. As open quantum systems (OQS), their properties are profoundly affected by coupling to the particle continuum \cite{okolowicz2003,Michel2009a,Betan2002,Michel2002,Sun2017,Johnson2020,Li2020,Ma2020Ne,MA20203N}. The Gamow-Berggren framework naturally incorporates continuum effects by replacing the standard harmonic-oscillator or real-energy basis with a complex-momentum Berggren basis, in which bound states, resonant Gamow states, and nonresonant continuum states are treated on equal footing through the Berggren completeness relation. Building on this framework, ab initio methods have successfully included continuum effects and computed resonances in light and medium-mass nuclei near the proton and neutron driplines, including the no-core Gamow shell model (NCGSM)~\cite{Papadimitriou2013,Li2019n4} continuum CC~\cite{Hagen2010f17,Hagen2012O,Hagen2012Ca}, Gamow IMSRG~\cite{Hu2019gimsrg}, and Gamow MBPT~\cite{Hu2020GSM,Zhang2022O,Zhang2023A16}.

Meanwhile, as the scope of ab initio nuclear theory has expanded to global calculations of heavy nuclei, predictions with quantified uncertainties have enabled significant progress in searches for BSM physics. Examples include calculations of neutrinoless double-beta decay matrix elements \cite{yao2020,belley2021,novario2021,belley2023,belley2024,Lian2026:0vbb}, precise beta decays \cite{glickmagid2022,longfellow2024,gennari2025}, muon-to-electron conversion \cite{haxton2023,heinz2024arXiv}, and nuclear Schiff moments \cite{engel2013,kia2025arXiv}. The next section focuses on one such advancement: DM direct detection.

\section{Dark Matter - Nucleus Scattering}

The nature of DM, invisible matter inferred only through its gravitational influence on other objects, is a major frontier of modern physics and astronomy \cite{RevModPhys.90.045002,doi:10.1146/annurev-astro-082708-101659,Undagoitia_2015,Bertone2018,Liu-NP13-212,DelNobile2022}. A leading candidate for DM is the weakly interacting massive particle (WIMP) \cite{engel1992,JUNGMAN1996195,Roszkowski_2018,Schumann_2019}, which is favored by many BSM theories and has an expected mass range from a few GeV to a few hundred TeV. Numerous direct-detection experiments worldwide aim to directly detect DM by observing the nuclear recoil from the scattering of galactic WIMPs off particular target nuclei. However, the lack of a convincing WIMP signal after decades of effort has increasingly motivated the search for a broader range of DM candidates, such as light DM (MeV-scale) models \cite{Trickle2022,Hochberg2022,Bell2024,Dutta2024,Dutta2025}. While a tremendous number of other DM models exist, this review will focus on the WIMP and light DM models.

To meaningfully interpret the results of direct-detection searches for WIMP or light DM, accurate theoretical DM-nucleus cross sections are required to predict the rate of potential DM scattering events \cite{DelNobile2022}. Such calculations are tremendously complicated and require a solid theoretical underpinning in both particle and nuclear physics: first to determine how WIMPs interact with individual nucleons in nuclei, then to fold in the relevant nuclear structure for detector nuclei \cite{hu2022b}. The final DM–nucleus cross section is expressed as a product of nuclear responses, also called nuclear structure factors or form factors, and the relevant coupling constants. Consequently, these nuclear responses are indispensable inputs for interpreting DM direct-detection experiments and are accessible only through theoretical calculations for nuclei spanning the light-, medium- and heavy-mass regions.

Currently, research at the intersection of nuclear physics, particle physics, and cosmology often relies on phenomenological approaches to compute nuclear responses, such as the large-scale shell model (LSSM) or Helm and Klein–Nystrand models \cite{Brown2001,Caurier2005,Takaharu2020,Lewin1996,Klein1999}. However, because the effective operators in these models are poorly constrained by data, such phenomenological approaches have intrinsic deficiencies for the precision required by modern experiments. Indeed, uncertainties associated with nuclear responses have become one of the the dominant sources of error in DM direct detection. It is therefore important to develop microscopic, accurate, and uncertainty-quantified nuclear models to describe these scattering processes.

\subsection{Elastic Scattering}

The main focus of direct detection searches is the elastic scattering of WIMPs off target nuclei. Interpreting the results of these experiments and setting exclusion limits requires theoretical calculations of the DM–nucleus interaction cross section. This generally involves two steps: (i) Within a given new-physics model, DM interacts with quark and gluon degrees of freedom, which need to be matched into hadronic-level currents. Converting DM–quark amplitudes into DM–nucleon amplitudes requires the values of the quark currents inside the nucleon, namely the nucleon matrix elements \cite{engel1992,hoferichter2015eft}. An elegant strategy employs chiral EFT via the external source method to decompose these quark-level nucleon matrix elements into nucleon form factors and meson exchange currents involving nucleon and pion degrees of freedom, organized systematically via the chiral expansion in a chosen power counting scheme \cite{Krebs2020,cirigliano2012wimp,hoferichter2015eft,deramo2015,hoferichter2016SD,hoferichter2020cevns}. The associated low-energy constants encode non-perturbative physics and can be determined phenomenologically or computed using lattice QCD \cite{beane2014lqcd,chang2018lqcd,davoudi2021lqcd,hoferichter2015eft,hoferichter2020cevns}.
Another powerful tool for constructing DM-nucleon interactions is non-relativistic EFT (NREFT) \cite{fan2010,fitzpatrick2013,anand2014}, which directly builds an EFT with non-relativistic nucleon and DM fields as the degrees of freedom. This approach generates the most general set of DM-nucleon interactions based solely on the Galilean invariance and momentum conservation, enabling bottom-up, model-independent analyses. While NREFT loses the connection to symmetries of QCD, it can be recovered by matching to the chiral EFT framework \cite{hoferichter2015eft,hoferichter2016SD}. (ii) Finally, the expectation values of these DM–nucleon currents are evaluated within the many-body wave functions of the target nuclei, yielding the nuclear responses that encode all the necessary nuclear physics information  \cite{engel1992}. Chiral EFT provides a systematic expansion and consistent treatment of both nuclear forces and one- and two-body currents (2BCs) of external probes coupling to nucleons, such as in WIMP-nucleus scattering~\cite{hoferichter2015eft,Krebs2020}. 
Indeed, 2BCs have a significant impact on both electroweak transitions in nuclei \cite{PhysRevLett.103.102502,PhysRevLett.122.029901,PhysRevLett.106.202502,PhysRevLett.107.062501,PhysRevC.87.035503,Bacca_2014,PhysRevLett.113.262504,PhysRevLett.117.082501,Gysbers2019} and DM scattering~\cite{PhysRevD.86.103511,PhysRevD.88.083516,PhysRevD.89.029901,PhysRevD.99.055031,PhysRevC.99.025501,PhysRevLett.122.071301}.

Scalar and axial-vector couplings are the dominant interactions for many WIMP candidates, giving rise to spin-indepentdent (SI) and spin-dependent (SD) cross sections, respectively \cite{engel1992,freese2013}. For SI searches, the leading nuclear responses are the Fourier transform of the proton and neutron density distributions within the nucleus. While proton distributions are well-constrained by electron scattering experiments, neutron distributions are poorly known due to extremely limited data~\cite{Ruso2025,Adhikari2021,hu2022a}. This leads to large uncertainties in constraining phenomenological model parameters. The neutron-skin thickness, $R_{\rm skin}$, defined as the difference between point neutron and proton radii, $R_{\rm n/p}$, is the most direct measure of the neutron distribution, so accurate theoretical calculations could significantly reduce uncertainties in phenomenological models. Fortunately, nuclear ab initio calculations have successfully described the $R_{\rm skin}$ \cite{hu2022a,Novario2023}. Calculations of the full nuclear responses from first principles are also ongoing, with results expected soon. 

\setcounter{figure}{2}
\setcounter{subfigure}{0}
\begin{subfigure}
\setcounter{figure}{2}
\setcounter{subfigure}{0}
    \centering
    \begin{minipage}[b]{0.8\textwidth}
        \includegraphics[width=14cm]{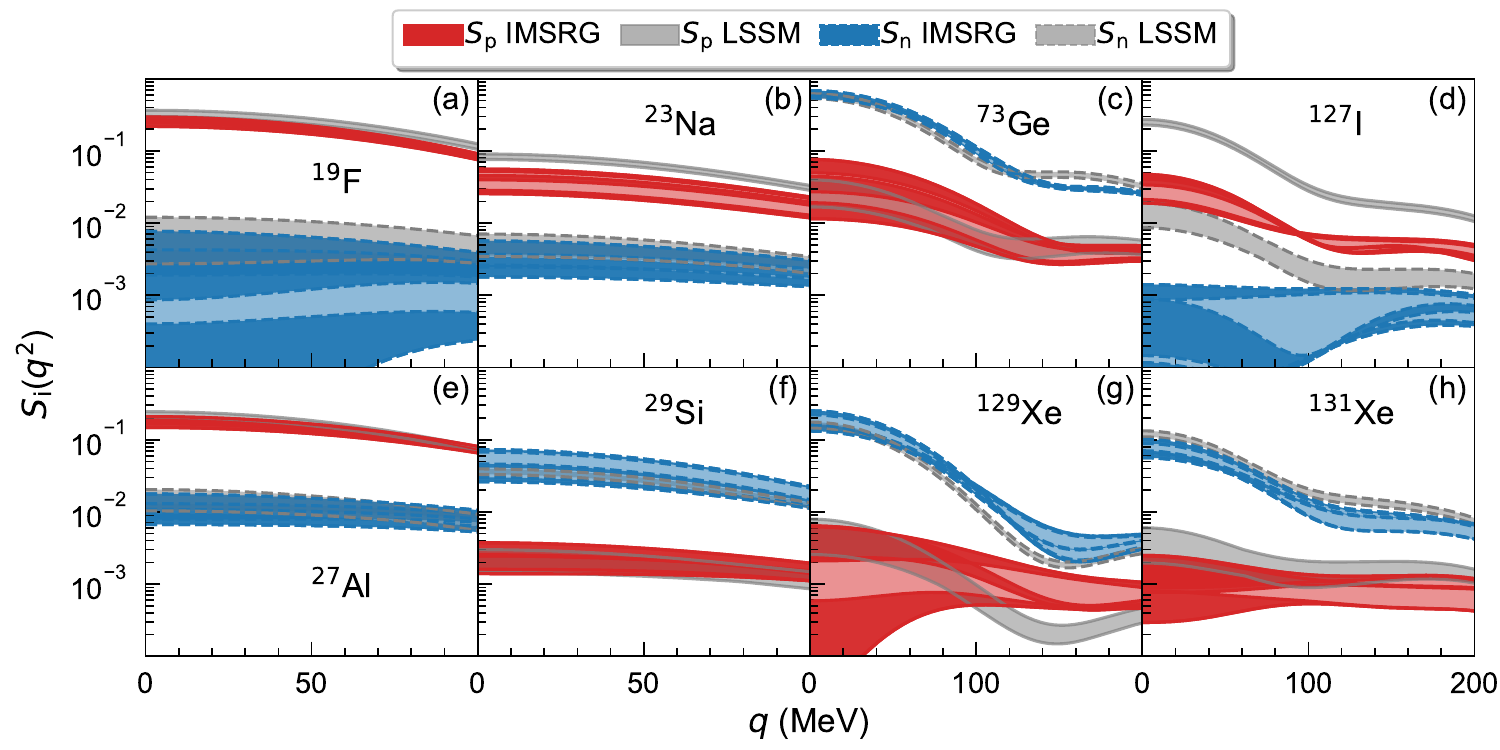}
        \caption{Ab initio axial-vector structure factors $S_i$ (include longitudinal and electric transverse multipoles) as a function of momentum transfer $q$. The gray bands indicate large-scale shell model (LSSM) calculations. The IMSRG bands depict ab initio results that spread in results from different interactions (lighter bands) and uncertainties in 2BCs (darker bands), while LSSM bands are from 2BC uncertainties only. Figure from Ref.~\cite{hu2022b}.}
        \label{fig:Subfigure 1}
    \end{minipage}  
   
\setcounter{figure}{2}
\setcounter{subfigure}{1}
    \begin{minipage}[b]{0.5\textwidth}
        \includegraphics[width=\linewidth]{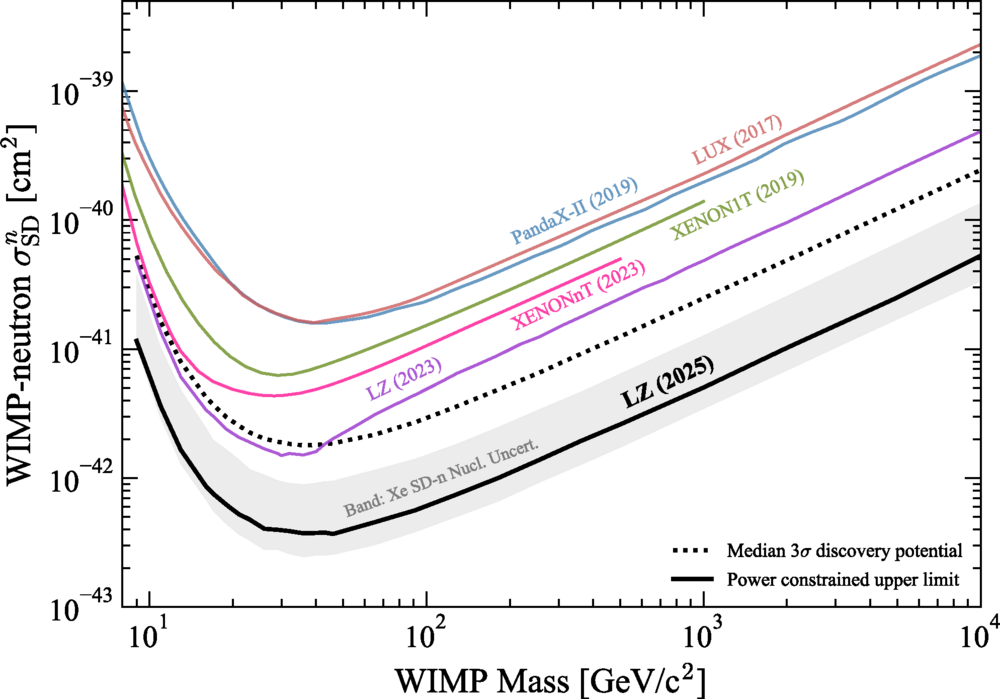}
        \caption{The 90\% confidence level exclusion limit on the spin-dependent WIMP-nucleon cross section from the LUX-ZEPLIN (LZ) experiment, compared with other experiments. The gray band indicates the uncertainty originating from the nuclear responses shown in panel (2a). Figure from Ref.~\cite{Aalbers2025lz}.} \label{fig:Subfigure 2}
    \end{minipage}

\setcounter{figure}{2}
\setcounter{subfigure}{-1}
    \caption{Ab initio structure factors and their impact on the LUX-ZEPLIN (LZ) dark-matter direct-detection experiment.}
    \label{fig:2}
\end{subfigure}

For SD searches, ab initio nuclear theory has achieved basis-space converged nuclear responses even in heavy nuclei by combining recent advances in natural orbitals with three-nucleon forces expressed in large model spaces. Fig.~\ref{fig:2} shows ab initio results for nuclear axial-vector structure factors $S_p$ (proton only) and $S_n$ (neutron only), compared with those from the LSSM. As an example, Fig.~\ref{fig:Subfigure 2} shows how the LUX-ZEPLIN (LZ) experiment uses these structure factors to derive its final exclusion plot. The gray band represents the uncertainty from the nuclear structure factors. We can see that uncertainty from nuclear physics remains a large component of the total error budget. This motivates further studies to reduce these nuclear physics uncertainties, for which the nuclear force is a leading contributor in ab initio calculations.

\subsection{Inelastic scattering}

Beyond elastic scattering, inelastic scattering provides an additional channel for DM detection. In this process, DM excites target nuclei to low-lying states, and the subsequent de-excitations yield visible signals. DM direct-detection experiments primarily focus on elastic scattering because it offers a simpler, more straightforward signature with higher event rates (cross section). 
While the inelastic channel lacks these advantages, it has received increasing attention, not only as a complementary search strategy but also for its unique sensitivity. This sensitivity arises because the ground and excited nuclear states have different structures, and the associated nuclear responses depend sensitively on the details of the DM-nucleon interaction. This allows for constraints on WIMP properties beyond what is possible with elastic scattering alone. In addition, the inelastic channel can provide better sensitivity than elastic scattering in certain regions of parameter space. Since the inelastic signal includes the nuclear de-excitation energy, it appears at higher visible energies, typically at the MeV scale, where neutrino-induced backgrounds are substantially lower than in the low-energy elastic-recoil region. As a result, although the inelastic scattering cross section is generally smaller than the elastic one, the reduced background can lead to better overall sensitivity. Inelastic WIMP scattering has been studied theoretically~\cite{Arcadi2019,Baudis2013}, and searched for experimentally by several collaborations~\cite{Suzuki2018XMASS,Aprile2021xenon1t}. 

Beyond WIMPs, inelastic scattering is also a powerful probe of light DM \cite{Dutta2022}. A recent approach proposes searching for beam-produced DM in fixed-target experiments, such as beam-dump neutrino experiments. This model shows that the inelastic channel can achieve significantly better reach than the elastic channel \cite{Dutta2023}. Another recent work finds that for probing cosmic-ray boosted DM at large-volume neutrino detectors, such as Borexino, DUNE, Super-K, Hyper-K, and JUNO, the inelastic channels generally provide better sensitivity than the elastic scattering for a large region of light DM parameter space \cite{Dutta2024}. These studies have so far relied on LSSM nuclear models \cite{Dutta2022,Dutta2023,Dutta2024,Brown2001,Caurier2005,Takaharu2020}, and it would be valuable to benchmark them with nuclear ab initio calculations \cite{Dutta2026sbnd}.

Accurate theoretical calculations of both nuclear excitation energies and transition matrix elements -- specifically the Coulomb, longitudinal, magnetic transverse, and electric transverse multipoles -- between the ground state and excited states are essential for interpreting inelastic-scattering searches. Due to this inherent sensitivity to the detailed nuclear structure of the excited states, inelastic scattering calculations currently introduce large theoretical uncertainties. Nuclear ab initio approaches are now poised to compute nuclear responses for inelastic scattering, as the underlying technology is the same as that used for the elastic case. A key difficulty, however, is that while elastic scattering only requires the ground-state wave function, inelastic calculations require computing wave functions for multiple excited states as well as the high-rank transition matrix elements between them. This significantly increases the computational cost, but the task is manageable with modern methods and high-performance computing resources. Work on ab initio calculations for these nuclear responses is ongoing and results are expected soon.

\section{THE FUTURE}

Ab initio nuclear theory continues to advance rapidly, driven by theoretical innovations, algorithmic development, and the growth of high-performance computing. The future of the field will likely be shaped by three interconnected endeavors: extending its reach to more complex nuclear structures and reactions, achieving higher precision with rigorous uncertainty quantification, and deepening the synergy with experimental frontiers.

\section{SUMMARY}

Ab initio nuclear theory has transitioned from light-nucleus benchmarks to a powerful predictive framework with controlled approximations, spanning light-, medium-, and heavy-mass systems. This review highlighted some recent achievements, including the landmark first-principles computation of $^{208}$Pb and its implications for the neutron skin and the nuclear-matter properties relevant to neutron stars. We also summarized the rapid advances in describing more complex systems like deformed nuclei with strong collective correlations and exotic isotopes near the driplines treated as open quantum systems. A further focus was the application of these developments to dark matter (DM) searches. We explained why accurate nuclear structure inputs are essential for interpreting experiments for both weakly interacting massive particle (WIMP) and light DM candidates, and summarized current challenges and recent successes in both elastic and inelastic scattering channels.

\section*{Conflict of Interest Statement}
%All financial, commercial or other relationships that might be perceived by the academic community as representing a potential conflict of interest must be disclosed. If no such relationship exists, authors will be asked to confirm the following statement: 

The authors declare that the research was conducted in the absence of any commercial or financial relationships that could be construed as a potential conflict of interest.

\section*{Author Contributions}
%The Author Contributions section is mandatory for all articles, including articles by sole authors. If an appropriate statement is not provided on submission, a standard one will be inserted during the production process. The Author Contributions statement must describe the contributions of individual authors referred to by their initials and, in doing so, all authors agree to be accountable for the content of the work. Please see  \href{https://www.frontiersin.org/about/policies-and-publication-ethics#AuthorshipAuthorResponsibilities}{here} for full authorship criteria.
The author confirms being the sole contributor of this work and has approved it for publication.

\section*{Funding}
This work was supported by the School of Physics and State Key Laboratory of Nuclear Physics and Technology at Peking University.

\section*{Acknowledgments}
%This is a short text to acknowledge the contributions of specific colleagues, institutions, or agencies that aided the efforts of the authors.
I thank the International Union of Pure and Applied Physics (IUPAP) C12 Commission for this opportunity. I am honored to be the recipient of the 2025 IUPAP Early Career Scientist Prize in Nuclear Physics (C12). I also gratefully acknowledge the Cyclotron Institute at Texas A\&M University for its support in recent years.

\bibliographystyle{Frontiers-Vancouver} % Many Frontiers journals use the numbered referencing system, to find the style and resources for the journal you are submitting to: https://zendesk.frontiersin.org/hc/en-us/articles/360017860337-Frontiers-Reference-Styles-by-Journal
\bibliography{Ref}

%%% Make sure to upload the bib file along with the tex file and PDF
%%% Please see the test.bib file for some examples of references

%%% If you don't add the figures in the LaTeX files, please upload them when submitting the article.
%%% Frontiers will add the figures at the end of the provisional pdf automatically
%%% The use of LaTeX coding to draw Diagrams/Figures/Structures should be avoided. They should be external callouts including graphics.

\end{document}